\documentclass[preprint2]{aastex}

\input{psfig.sty}                              

\shorttitle{UY Camelopardalis: an analogue of SX Phe and RR Lyr-type variables}
\shortauthors{Zhou \& Liu}

\begin{document}

\title{Monoperiodic $\delta$ Scuti star UY Camelopardalis:
an analogue of SX Phe and RR Lyr-type variables}

\author{A.-Y. Zhou and Z.-L. Liu}
\affil{National Astronomical Observatories, Chinese Academy of Sciences,
      Beijing 100012, China.}

\email{aiying@bao.ac.cn}

\begin{abstract}
We present the results of a four-year photometric study of
the high-amplitude $\delta$~Scuti star (HADS) UY Camelopardalis.
Analyses on the available data from 1985 to 2003 show that UY Cam is monoperiodic.
Fourier solutions for individual data sets do not reveal
period changes in the star.
Although forced parabolic fits to the $O-C$ residuals indicate measurable period change,
the distribution of the data points in
$O-C$ diagram and the deviations between fits and observations suggest the period
change is still not established.
We evidence the presence of cycle-to-cycle and longer time-scale amplitude variations.
The pulsation amplitude seemed to change from 1985 to 2000s,
but it kept constant in 2000--2003.
UY Cam locates in the upper portion of the instability region of $\delta$ Sct variables.
Photometric properties and estimated physical parameters reveal that UY Cam is
an interesting object concerning its poorer metallicity,
longer period, higher luminosity, lower surface gravity and larger radius among the HADS.
UY Cam could be a younger (age=0.7$\pm$0.1\,Gyr) Population~I HADS with
poor metal abundances (Z=0.004) evolving on
its post main sequence shell hydrogen-burning evolutionary phase.
UY Cam intervenes among the Pop.~I/II HADS and RRc Lyr variables.
These characteristics suggest the star to be an analogue of HADS,
SX Phe and RRc Lyr-type variables.
\end{abstract}


\section{Introduction}
$\delta$ Scuti stars are regularly pulsating variables situated in
the lower classical Cepheid instability strip on or near the main sequence (MS).
In general, the period range of $\delta$ Sct stars lies between
0.02 and 0.25\,d and the spectral types
range from A2 to F2.
The majority of $\delta$ Sct stars pulsate with a number of
non-radial $p$-modes simultaneously excited to low amplitudes,
but some are (pure) radial pulsators with larger amplitudes and
others pulsate in a mixture of radial and nonradial modes.
We launched a mission dedicated to the investigations of poorly-studied $\delta$ Sct stars
in 1996.
UY Camelopardalis (=HIP 39009=GSC 04369-01129, $\alpha_{2000} = 07^{\rm h}58^{\rm m}59^{\rm s}.0$,
$\delta_{2000} = +72^{\circ}47'23^{\prime\prime}.7$, $V$=11.44\,mag, A3-A6~III;
\citealt*{rodr00})
belongs to high amplitude $\delta$ Sct stars (HADS)
subclass with less knowledge on its nature. Therefore it was selected as
one of the targets for the mission.

UY Camelopardalis was discovered to be a variable by \citet{bake37},
who classified the star as a Cepheid. Observations on five nights in 1962 led
\citet{will64} to resolve a period of 0.267 days. Williams suggested UY Cam as an RRc Lyrae star.
He pointed out that further observations with a larger
telescope (than his 20-inch one) are necessary to establish the possible existence
of cycle-to-cycle changes in the light curves and to study their details.
During 1946 and 1965, \citet{beye66} collected 21 maxima and derived a constant period.
According to the author, the light curve was unstable and its amplitudes changed
from 0.17 to 0.50\,mag.
However, the potential instability of the light curves of UY Cam, suspected by
both Williams and Beyer, was not observed late in 1985 by \citet{brog92}, who obtained a total of
five nights Johnson $BV$ data covering five maxima.

Considering the insufficiency of previous data in revealing the possibility of amplitude
variations, we observed UY Cam from 1999 to 2003 with an emphasis of time-resolving and
longer time baseline. We collected a number of CCD and photoelectric
photometric data in the Johnson $V$ band.
In this paper, we present the results of a comprehensive analysis aimed at
short and/or secular behaviour of light variations using all available data.
Section~2 contains an outline of the observational journal and data reduction.
Section~3 devotes to analyses of the data. The nature of the variable is
briefly discussed in Section~4 and our main results are summarized in Section~5.

\section{Data acquisition}
\label{sect:Obs}
New observations of UY Cam were secured
between 1999 November 17 and 2003 March 3.
The data consist of 8337 measurements (merged bins in 60-s) in Johnson $V$ band (129.5\,h)
collected in 19 observing nights.
A journal of the observations is given in Table~\ref{Tab:Obs-log}.

\subsection{CCD photometry}
From 1999 November 19 to 2000 January 24,
Johnson $V$ photometry of UY Cam was performed with
the light-curve survey CCD photometer \citep*{wei90,zhou01a}
mounted on the 85-cm Cassegrain telescope at the Xinglong Station of
the Beijing Astronomical Observatory (BAO) of China.
The photometer employed a red-sensitive Thomson TH7882 576$\times$384 CCD
with a whole imaging size of 13.25$\times$8.83~mm$^{2}$
corresponding to a sky field of view of $11'.5\times7'.7$, which allows
sufficient stars to be toggled in a frame as reference.
Depending on the nightly seeing the integration times varied from 20 to 60\,s.
All the monitored reference stars in the field of UY Cam were detected as nonvariables
within the observational error ($\sim$0.006\,mag),
which is the typical standard deviation of magnitude differences between the reference stars.
Among them,
GSC 4369-1457 (
$\alpha_{2000} = 07^{\rm h}58^{\rm m}11^{\rm s}.12$,
$\delta_{2000} = +72^{\circ}45'51^{\prime\prime}.8$, $V$=11.6\,mag)
was confirmed to be the best comparison against the variable.
Then the differential magnitudes of UY Cam were measured with respect to GSC 4369-1475.
Atmospheric extinction was ignored because of the proximity between the
two stars.
The procedures of standard data reduction, including on-line bias subtraction,
dark reduction and flatfield correction, were outlined in \citet{zhou01a}.
%
%
\begin{table}[t]
  \caption[]{Journal of Johnson $V$ photoelectric and CCD photometry of UY Cam.
  The JD is in 2450000+ days.}
  \label{Tab:Obs-log}
  \begin{center}\begin{tabular}{ccccl}
\tableline\tableline
Run    & Night(UT)   & JD   & \hspace{-5pt} Time interval(d) & \hspace{-6pt}Points\\
\tableline
2000   & 2000.01.07  &1551  & 0.126  & 273  \\
       & 2000.01.13  &1557  & 0.291  & 498  \\
       & 2000.01.14  &1558  & 0.157  & 350  \\
       & 2000.01.15  &1559  & 0.261  & 469  \\
       & 2000.01.16  &1560  & 0.333  & 567  \\
       & 2000.01.17  &1561  & 0.317  & 601  \\
       & 2000.01.19  &1563  & 0.268  & 386  \\
       & 2000.01.23  &1567  & 0.089  & 161  \\
       & 2000.01.24  &1568  & 0.310  & 430  \\
2002   & 2002.01.23  &2298  & 0.309  & 438  \\
       & 2002.01.26  &2301  & 0.352  & 491  \\
       & 2002.01.27  &2302  & 0.362  & 515  \\
       & 2002.01.28  &2303  & 0.379  & 524  \\
       & 2002.01.29  &2304  & 0.364  & 507  \\
       & 2002.01.31  &2306  & 0.203  & 289  \\
2003   & 2002.12.31  &2640  & 0.337  & 486  \\
       & 2003.02.28  &2699  & 0.356  & 514  \\
       & 2003.03.02  &2701  & 0.369  & 532  \\
       & 2003.03.03  &2702  & 0.212  & 306  \\
\tableline
  \end{tabular}
  \end{center}
\end{table}

\subsection{Photoelectric photometry}
From 2002 January 23 to 2003 March 3, UY Cam was reobserved
with the three-channel high-speed photoelectric photometer \citep{jian98}
attached to the same telescope. This detector is used in
the Whole Earth Telescope campaigns (WET; \citealt{nath90}).
GSC 4380-1705 ($\alpha_{2000} = 08^{\rm h}00^{\rm m}13^{\rm s}.29$,
$\delta_{2000} = +72^{\circ}43'47^{\prime\prime}.9$, $V$=11.2\,mag) was
chosen as the comparison star.
The variable, comparison star and sky background were simultaneously monitored in
continuous 10-s intervals through a standard Johnson $V$ filter throughout
the two observing runs in 2002 and 2003.
The typical observational accuracy with the three-channel photometer is about
0.006\,mag.

\section{Data analysis}
\label{sect:FA}

\subsection{Frequency detection}
Previous works on UY Cam did not suggest any other additional pulsation frequencies
except the known primary frequency.
We decided to make a further detection to find a complete set of pulsation frequencies.
To do so, we merged all the data collected from 1999 to 2003.
The frequency analysis was carried out by using the programmes {\sc period98}
\citep{breg90, sperl98}
and {\sc mfa} \citep{hao91, liu95},
where single-frequency Fourier transforms and multifrequency least-squares fits were processed.
The two programmes use the Discrete Fourier Transform method \citep{deem75}
and basically lead to identical results.
We first computed the noise levels at each frequency using the residuals
at the original measurements with all trial frequencies prewhitened.
Then the confidence levels of the trial frequencies were
estimated following \citet{scar82}.
Last, to judge whether a peak is or not significant in the amplitude spectra
we followed the empirical criterion of \citet{breg93},
that an amplitude signal-to-noise (S/N) ratio larger than 4.0 usually
corresponds to an intrinsic peak of the variable.
Note that the S/N criterion assumes a good spectral window typical of
multisite campaigns.
However, in the case of single-site observations, the noise level can be
enhanced by the spectral window patterns of the noise peaks and possible
additional frequencies. Therefore a significant peak's S/N value might be
a little less than 4.0 in single-site case.

%
\begin{figure}[ht]
   \vspace{2mm}
   \hspace{-5mm}\psfig{figure=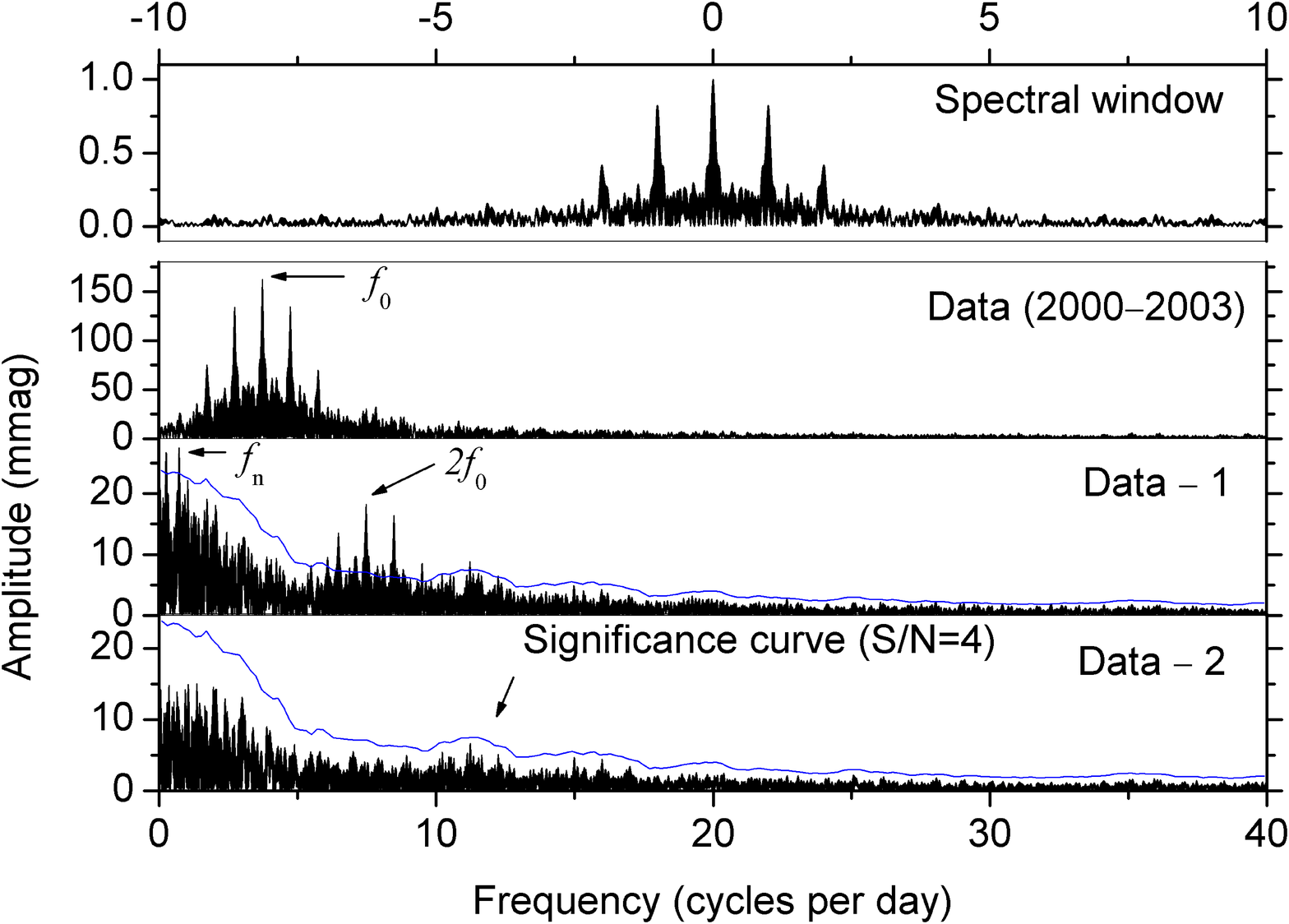,width=85mm,height=90mm,angle=-0.0}
   \caption{The spectral window and amplitude spectra of UY Cam (2000--2003).
   Note the high noise in 0--5 d$^{-1}$ is visible
   in last two panels, where significance curves with S/N=4.0 were drawn. }
   \label{Fig:power}
\end{figure}

\begin{figure}
   \vspace{2mm}
   \hspace{-5mm}\psfig{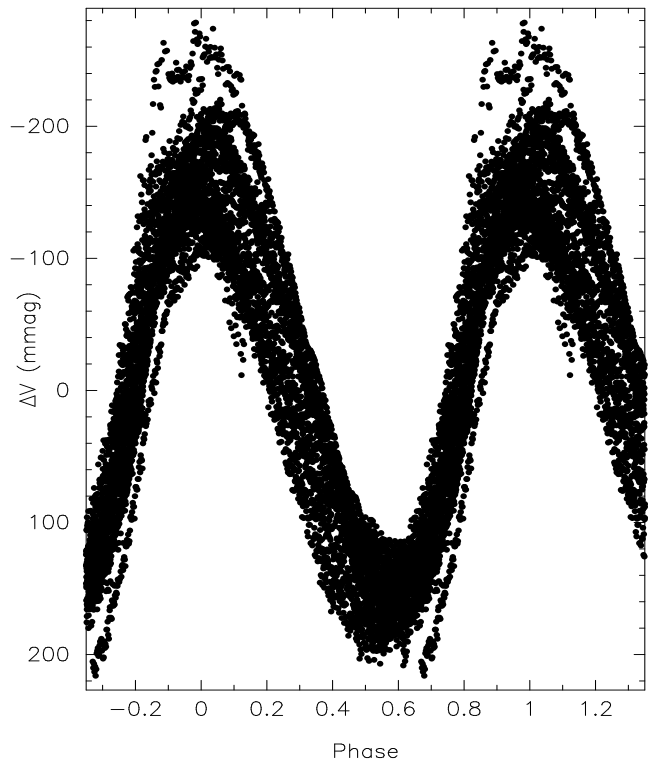}
   \caption{Phase diagram of the observed differential light curves of UY Cam
   in 2000--2003. Phases were calculated relative to HJD 2451551.38219 at which
   phase got zero.}
   \label{Fig:phase-diagram}
\end{figure}

\begin{figure*}
   \vspace{2mm}
   \hspace{4mm}\psfig{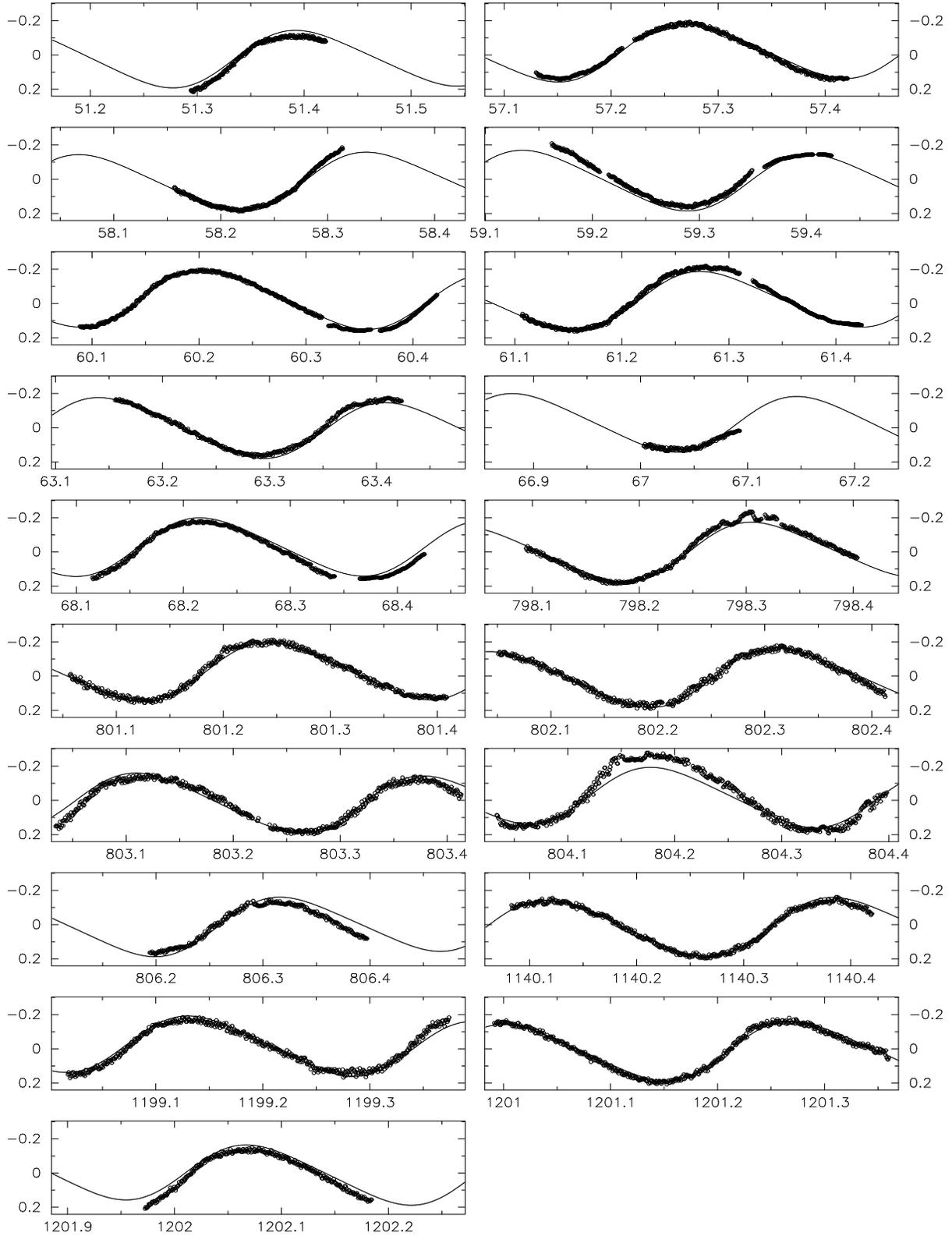}
   \caption{Observed differential light curves of UY Cam in 2000--2003
   together with the sinusoids (solid lines, with $f_0$=3.7447 d$^{-1}$ and $2f_0$).
   Abscissa in HJD 2451500+ days, ordinate in magnitude.}
   \label{Fig:lightcurve}
\end{figure*}

For the judgement of significant peaks the amplitude spectra and spectral window
are given in Fig.~\ref{Fig:power}, in which each spectrum panel corresponds to
the residuals with all the previous frequencies prewhitened.
The last panel (`Data -- 2f') shows the residuals after subtracting the fit
of the two outstanding frequencies, together with the significance
curve -- spline-connected points -- four times the noise points obtained in
a spacing of 0.2 d$^{-1}$.
The observational noise is frequency-dependent and it was defined as
the average amplitude of the residuals in a range of 1\,d$^{-1}$ (11.57 $\mu$Hz),
close to the frequency under consideration.
We see high noise in the low frequency region (0--5 d$^{-1}$), which might be caused by
instrument shifts and nonidentical zero points of different runs or
even night-to-night zero point shifts.
Apart from the outstanding primary frequency $f_0$ and its first harmonic $2f_0$,
a peak at 0.7394\,d$^{-1}$ ($f_n$) with S/N=5.0 has an effect on the spectrum (see
the last two panels of Fig.~\ref{Fig:power}).
Frequencies $f_0$ and $2f_0$ fit the light curves with a standard deviation of residuals
of $\sigma$=0.0312\,mag and a zero point $-$0.0019\,mag. If $f_n$ was considered,
the values become to be 0.0242 and 0.00164\,mag, respectively. The term $f_n$ improved the fitting quality
evidently by about 22\%.
However, there is no reason allowing us to attribute this frequency ($f_n$) to the variable,
we regard it as a noise content involved in the data.
So prewhitening $f_n$ played a role of denoising the data.
As a final result, the pulsation frequency $f_0$=3.7447\,d$^{-1}$ and
its harmonic $2f_0$ with amplitudes of 0.1638 and 0.020\,mag, respectively,
are detected to be intrinsic to UY Cam.
No additional peak is significant in the residual spectrum.
Therefore, the variable is monoperiodic.
Figure~\ref{Fig:phase-diagram} displays the light curves folded with the main frequency $f_0$.
The differential light curves along with the fits using
$f_0$ and $2f_0$ after denoising are presented in Fig.~\ref{Fig:lightcurve}.
It is noted that there exists a zero point shift on one night in the phase diagram.
This figure appears to be an evidence of amplitude variability.
We will explore this in more detail in next subsection.
%
%
\begin{table*}[t]
   \caption[ ]{Results of the frequency analysis of UY Cam based on the data in 2000--2003.
   Epoch in HJD 2451550.0+ days. Calculated errors are given below the quantities. }
   \label{Tab:freq}
   \begin{center}\begin{tabular}{crrrlrc}
\tableline\tableline  
\multicolumn{2}{c}{ Freq.}
                    &Ampl.  &  Phase& Epoch & S/N & Conf.\\
\multicolumn{2}{c}{( d$^{-1}$)}
                    &(mmag) &  (0--1) & (days) &     & (\%) \\
\tableline\noalign{\smallskip}
 $f_{0}$&  3.74475 & 163.8 &  0.266& 7.010 &44.3 & 100  \\
        & $\pm$.00035 & ~~$\pm$1.9 &  ~$\pm$.010& ~~~~~ &     &      \\
2$f_{0}$&  7.48685 &  20.0 &  0.188& 7.108 &11.9 & 100  \\
        &   $\pm$.00270 & ~~$\pm$1.9 &  ~$\pm$.095& ~~~~~ &     &      \\
\tableline
   \end{tabular}
   \end{center}
\end{table*}

Fourier parameters of the best-fitting sinusoids,
$ m(t) = A_0 + \sum_{i=1}^{N}{A_i \cos(2\pi f_i t + \phi_i)}$,
are listed in Table~\ref{Tab:freq}, where the errors in frequencies, amplitudes and
phases were estimated through the formulae of \citet{mont99}.
We assumed the root-mean-square deviation of the
observational noise to be of 0.00242\,mag, the standard deviation of the fit with
$f_0$ and $2f_0$ after removing the noise content ($f_n$), five times the practical observational accuracy.
We further used the real nights
with data for the time baseline rather than the span of observations.
Even in this way, we found the frequency errors might have been underestimated.
Because the single-frequency Fourier transform showed the second frequency term
at 7.48685\,d$^{-1}$ differing from $2f_0$ by 0.0026\,d$^{-1}$, which is larger than
the theoretically calculated errors
above (0.00007 and 0.00054\,d$^{-1}$ for $f_0$ and $2f_0$, respectively).
So we finally adopted the values five times the calculated errors so that
the error for $2f_0$ becomes 0.0027\,mag conforming with the difference 0.0026\,mag.
As mentioned by \citet{mont99}, this is a perfectly valid thing to do so,
since hidden correlations in the errors in the data lead to an underestimate of the true errors
because correlations in the noise can modify the calculating relations.

%

%

%
%
\begin{table*}[th]
  \caption[]{Subsets of data used in the time-dependent analysis.  }
  \label{Tab:dataset}
  \begin{center}
  \begin{tabular}{cccc}
\tableline\tableline
Run    & Date                     &Nights   & Measurements     \\
\tableline
1985   & 1985 Apr.14--28          & 5       & ~364$^{\star}$ \\

2000   & 2000 Jan.7--24           & 9       & 3735 \\

2002   & 2002 Jan.23--31          & 6       & 2786 \\

2003   & 2002 Dec.31--2003 Mar.3  & 4       & 1838 \\
\tableline
  \end{tabular}
  \end{center}
\begin{list}{}{}
\item[$^{\star}$] From Broglia \& Conconi, 1992, IBVS, No.3748
\end{list}
\end{table*}

\subsection{Frequency and amplitude variability}
It is possible to search the data for amplitude and/or frequency variability for the star.
We dealt with the data in individual subsets corresponding to different observing runs.
The CCD data in 1999 were ignored for their bad quality.
In addition, data set 1985 (5 nights) was adopted from \citet{brog92}.
Table~\ref{Tab:dataset} summarizes these data sets.
Individual Fourier analyses were then carried out for each set of the data.
However, we are aware that the Fourier results highly depend on the data structure,
i.e. the number of data points, sampling rate, time span, etc.
Fourier analysis assumes the frequency and amplitude of a signal
to be constant in the time domain. Therefore, our results for the time-dependence of
frequency and amplitude should be regarded as the averaged values
during the investigated period.

Given the trial frequency values of $f_0$=3.7447 d$^{-1}$ and its two harmonics
$2f_0$ and $3f_0$,
non-linear least-squares sinusoidal fittings to each data set were performed.
Our results are listed in Table~\ref{Tab:freq-var},
where the errors were estimated following the methods in previous section.
The standard deviations of fits for the four subsets are
$\sigma$=0.0117, 0.0222, 0.0382 and 0.0216\,mag, respectively.
The fitting error for data set 1985 agrees with its observational accuracy.
However, the fitting quality is generally low for the present data.
For sets 2000 and 2003, the errors are about four times
the observational accuracy,  while more than 6 times for set 2002.
In view of these inconsistent large fitting errors, the pulsation frequency and amplitude
seemed to change. However, from Table~\ref{Tab:freq-var},
within the error bars, the primary frequency were quite stable over the past years.
On the other hand, except the data set 1985, the amplitudes were basically
consistent with each other.

\begin{figure}[t]
   \vspace{2mm}
\plotone{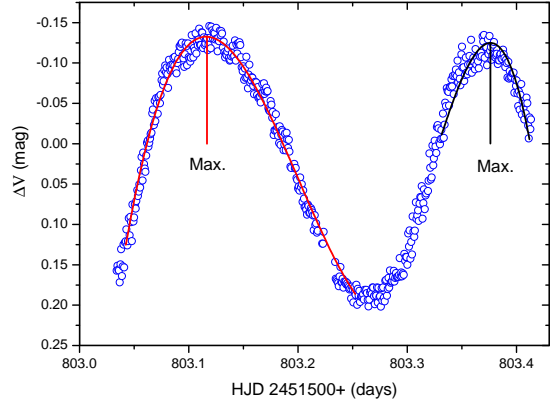}
   \caption{The light curve of UY Cam on 2002 January 28. The 3-order polynomial
   fittings to the different portions around the maxima show the determination of
   times of maxima. This figure also displays the cycle-to-cycle amplitude variation.  }
   \label{Fig:curve-020128}
\end{figure}

According to Figs.~\ref{Fig:phase-diagram} and \ref{Fig:curve-020128},
as well as the light curves on HJD 24512304 (2002 Jan.29,
the panel with abscissa '804' in Fig.~\ref{Fig:lightcurve}),
amplitude variation, at least at cycle-to-cycle level might present.
A glance at Table~\ref{Tab:freq-var} shows that
pulsation amplitudes in 2000--2003 were the same within the errors range.
However, the amplitude in 1985 was a little bit higher than the others.
Because the best fittings make all three Fourier parameters optimized,
we attempted to check the case when frequencies are fixed.
By fixing the frequency $f_{0}$=3.7447 d$^{-1}$ and its harmonics
$2f_{0}$= 7.4894 d$^{-1}$ and $3f_{0}$= 11.2341 d$^{-1}$ and
allowing their amplitudes and phases to change,
we obtained the amplitudes of the three frequencies for different years' data.
Deviations between fits and light curves are relatively smaller in 1985 data set
than in other three sets: 0.0149, 0.0226, 0.0383 and 0.0219\,mag for the four runs, respectively.
The results in Table~\ref{Tab:ampl-var} are consistent with those in Table~\ref{Tab:freq-var}.
Amplitudes for 2000, 2002 and 2003 were consistent with each other.
However, in both Tables~\ref{Tab:freq-var} and \ref{Tab:ampl-var}, the amplitude of
$f_0$ in 1985 is obviously higher than those in other three years.
In the estimated errors range, this difference probably is real.
Consequently, we think that the amplitude of $f_0$ changed in 1985, but it kept constancy
during 2000 and 2003.
In addition, the contribution of the third-order harmonic $3f_0$ to the light variations
could be neglected.
The picture of amplitude variation likes that mentioned
by \citet{breg00} that the HADS typically have less amplitude variation than those
low-amplitude $\delta$ Sct stars.
%
%
\begin{table*}[t]
  \caption[]{Fourier parameters of the best-fitting sinusoids for the
four subsets of data from 1985 to 2003.  }
  \label{Tab:freq-var}
  \begin{center}
  \begin{tabular}{crcc}
\tableline\tableline
Run   & \multicolumn{1}{c}{Frequency(d$^{-1}$)}
      & \multicolumn{1}{c}{Amplitude(mmag)}      & \multicolumn{1}{c}{Phase(0--1)}   \\
\tableline
1985  &$f_0$  ~3.745$\pm$.004 &   182.3$\pm$4.5 & .620$\pm$.015 \\
      &$2f_0$ ~7.492$\pm$.008 & \,~20.8$\pm$4.5 & .521$\pm$.126 \\
      &$3f_0$ 11.245$\pm$.015 & \,~~5.8$\pm$4.5 & .460$\pm$.134 \\
2000  &$f_0$  ~3.748$\pm$.001 &   166.8$\pm$2.4 & .077$\pm$.015 \\
      &$2f_0$ ~7.477$\pm$.009 & \,~16.1$\pm$2.4 & .825$\pm$.126 \\
      &$3f_0$ 11.233$\pm$.019 & \,~10.3$\pm$2.4 & .834$\pm$.106 \\
2002  &$f_0$  ~3.745$\pm$.002 &   165.2$\pm$3.0 & .328$\pm$.018 \\
      &$2f_0$~ 7.492$\pm$.012 & \,~21.9$\pm$3.0 & .896$\pm$.150 \\
      &$3f_0$ 11.284$\pm$.066 & \,~~6.5$\pm$3.0 & .501$\pm$.242 \\
2003  &$f_0$  ~3.745$\pm$.012 &   162.4$\pm$8.7 & .137$\pm$.090 \\
      &$2f_0$~ 7.489$\pm$.039 & \,~19.2$\pm$8.7 & .907$\pm$.280 \\
      &$3f_0$ 11.233$\pm$.150 & \,~~7.4$\pm$8.7 & .570$\pm$.358 \\
\tableline
  \end{tabular}
  \end{center}
\end{table*}


In order to further examine the stability of the primary frequency, we determined
the times of maximum light and made use of those existing maxima.
To determine a maximum, we applied a polynomial fit to
a portion of the light curve around each maximum and
took the extrema of the polynomials as the observed times of maximum light.
We usually tried several times of fitting by selecting polynomials
of different orders (e.g. 3-order, sometimes 5-order) and different regions
until a satisfactory fit was reached.
Generally, we choose a range with an amplitude of about one-third of the full amplitude.
See an illustration in Fig.~\ref{Fig:curve-020128}.
The fitting errors are typically 0.008\,mag, being in agreement with the observational precision.
Thus 16 maxima were determined from the present data.
The errors of maxima determination are about 0.00045\,d.
In addition,
in order to avoid the difficulties of fitting a polynomial to determine
the precise time of maximum of each peak, one could perform a sine fit including the
harmonics for the given season of observations, i.e. the observations
in 1985 and in 2000--2003. However, given that the relative amplitude of the harmonics
may change with respect to the amplitude of the fundamental frequency $f_0$
(and that this would change when the time of maximum occurs even if the frequency $f_0$
is absolutely constant), it may be safer to look at the time of maximum of
only the $f_0$ component, not the $2f_0$ or $3f_0$ components.
This way, we selected four times of maximum light for seasons
of 1985, 2000, 2002 and 2003: 24546173.4361, 2451560.21625, 2452301.2504 and 2452699.13721.
Their corresponding $O-C$ values are $-$0.04249, 0.00678, 0.00493 and 0.00214\,d.
It is noted that these times of maximum have a lag ranging from 0.00375 to 0.01197\,d
with respect to their observed peaks determined by polynomial fits.
Moreover, the fits with only $f_0$ is poor, so these times act as mean values.
%
%
\begin{table}[t]
  \caption[]{Pulsation amplitudes of the primary frequency ($f_0$=3.7447\,d$^{-1}$)
  of UY Cam and its two harmonics in 1985--2003.   }
  \label{Tab:ampl-var}
  \begin{center}
  \begin{tabular}{rrrrr}
\tableline\tableline
       &    \multicolumn{4}{c}{Amplitude (mmag)}         \\
       &    \multicolumn{1}{c}{1985}     &   \multicolumn{1}{c}{2000}
       &    \multicolumn{1}{c}{2002}     &   \multicolumn{1}{c}{2003}       \\
\tableline
$~f_0$ & 181.0$\pm$4.5 & 166.2$\pm$2.4 & 165.5$\pm$3.0 &162.1$\pm$8.7 \\
$2f_0$ & ~21.2$\pm$4.5 & ~14.9$\pm$2.4 & ~21.5$\pm$3.0 &~19.0$\pm$8.7 \\
$3f_0$ & ~~6.0$\pm$4.5 & ~10.2$\pm$2.4 & ~~6.3$\pm$3.0 &~~7.2$\pm$8.7 \\
\tableline
  \end{tabular}
  \end{center}
\end{table}

\begin{figure}[h]
   \vspace{2mm}
\plotone{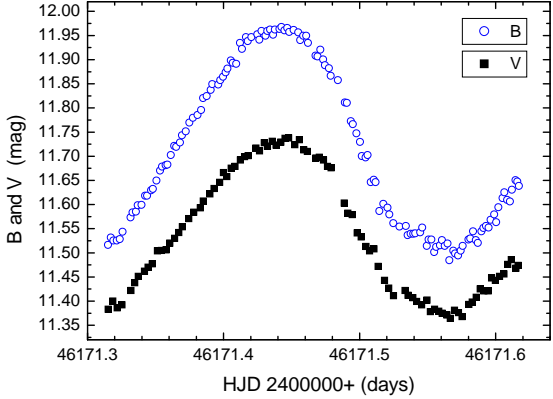}
   \caption{The $BV$ light curves of UY Cam on 1985 April 15. The phase difference between
   $B$ and $V$ colors can be ignored regarding the error of maximum determination.
   Data adopted from \citet{brog92}.  }
   \label{Fig:curve-BV}
\end{figure}

\begin{figure}[th]
   \vspace{2mm}
\plotone{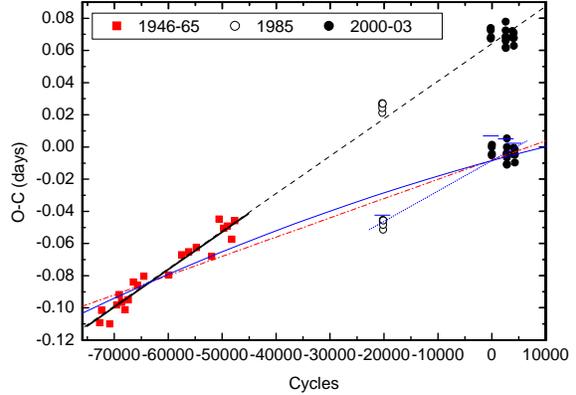}
   \caption{The $O-C$ diagram for UY Cam. Bars: season's mean values;
   bold solid line: linear, 1946--65 data;
   dot line: linear, 1985--2003 data;
   dash dot line: linear, all data;
   thin solid line: quadratic, all data.
   dash line: linear, all data but the 1985--2003 data were shifted by 0.07243\,d.
   The validity of this offset is unresovled.    }
   \label{Fig:O-C}
\end{figure}

We adopted five maxima from \citet{brog92}.
In addition, there are 21 maxima in $B$ band by \citet{beye66} in the literature.
In terms of the $BV$ light curves (IAU archives No.246E) of \citet{brog92},
we found the phase difference between $V$ and $B$ filters can be ignored.
In fact, when fixing $f_0$ and $2f_0$, fittings to the $BV$ data produced an equivalent
phase difference of 0.0003 days, which is less than the errors of maxima determination.
We refer the reader to Fig.~\ref{Fig:curve-BV}, where the $BV$ light curves on
HJD 2446171 were drawn.
So we combined the 21 maxima in $B$ band with the others in $V$.
In total, we have 42 observed times of maximum light (see Table~\ref{Tab:O-C}).
Our first maximum was taken as the initial epoch and $P_0$=1/$f_0$=0.26704\,d as the trial period,
i.e. we defined an initial ephemeris
\begin{equation}
{\rm HJD_{max}} = 2451557.27203 + 0.26704 E .
\label{eq:epoch}
\end{equation}
Then the observed minus the calculated times of maxima ($O-C$ residuals) and
the cycles elapsed from this epoch were calculated.
We noted the counting of cycles are integers so whenever a maximum occurred at
a time near to next cycle, i.e. more than half a cycle away from the previous cycle,
the next cycle number was chosen for this maximum.
To determine whether the times of maximum light are consistent with a constant period
($O-C$ residuals lie on a truly straight line) or a changing period (on a parabola),
a parabolic fit was applied to the data. We fit the $O-C$ residuals to the equation
$O-C = \Delta T_0 + \Delta P\,E + 0.5\beta E^{2}$,
following e.g. \citet{kepler00} and \citet{zay01}.
Where $\Delta T_0 = T_0^{\rm new} - T_0^{\rm ini}$ (improved new epoch minus assumed
initial epoch), $\Delta P$ refers to improved minus adopted trial period,
and $\beta$ refers to $\frac{{\rm d}P}{{\rm d}t}$, period change rate.
Using all available times of maximum light, we obtain
\begin{equation}
  \begin{array}{rll}
  O-C =  -0.00853~~    + & 9.11\cdot 10^{-7} E & - ~ 4.43\cdot10^{-12} E^{2}\\
      \pm 0.00234~~   \pm& 2.21                &\pm~ 3.38       \\
  \end{array}
\label{eq:LQfit}
\end{equation}
with a fitting error of $\sigma$=0.0104\,d.
If the quadratic term was ignored --- disregarding period change,
fit becomes to be linear as
\begin{equation}
  \begin{array}{rcl}
  O-C =  -0.00819  & + & 1.19199\times 10^{-6} E\\
       \pm0.00234  &\pm& 0.05309\nonumber\
 \label{eq:Lfit}
\end{array}
\end{equation}
with  $\sigma$=0.01049\,d (see the dash-dot line in Fig.~\ref{Fig:O-C}).
If the (5+16) maxima in 1985--2003 are replaced by their four corresponding mean times
(plotted as lengthened bars in Fig.~\ref{Fig:O-C}),
parabolic fitting to the 25 data points (21 points from 1946--65) leads us to
\begin{equation}
  \begin{array}{rl}
  O-C =  -0.0000180187   & + ~ 4.58639\times 10^{-7} E \\
       \pm0.00311~~~~~~~~   &\pm~ 2.54803                \\
   &  - ~1.3416\times 10^{-11}  E^{2}\\
   & \pm~0.3528
  \end{array}
\label{eq:LQfit-mean}
\end{equation}
($\sigma$=0.00829\,d).
Compared with the fits in Eq.(\ref{eq:LQfit}) and (\ref{eq:Lfit}),
the quality of fitting was improved in Eq.(\ref{eq:LQfit-mean}).
Equation (\ref{eq:LQfit}) means a rate of period change (decrease)
$\frac{{\rm d}P}{{\rm d}t} = (-8.86\pm6.76)\times 10^{-12}$~d cycle$^{-1}$ =
$(-3.32\pm2.53)\times 10^{-11}$~d\,d$^{-1}$,
and $\frac{1}{P}\frac{{\rm d}P}{{\rm d}t}=(-4.54\pm3.46)\times 10^{-8}~{\rm yr}^{-1}$.
In the case of Eq.(\ref{eq:LQfit-mean}),
$\frac{{\rm d}P}{{\rm d}t} = (-10.0\pm2.6)\times 10^{-11}$~d\,d$^{-1}$ and
$\frac{1}{P}\frac{{\rm d}P}{{\rm d}t}=(-13.68\pm3.56)\times 10^{-8}~{\rm yr}^{-1}$,
is about three times the previous one.
The values of period change are comparable with those listed in
\citet{breger98} and \citet{rodr95}.
We note here that error bars for the maxima from 1985 to 2003 are around 0.0005 days,
which are about a quarter of the size of the symbols in Fig.~\ref{Fig:O-C},
so they are invisible in current figure.
%
%
%

At a first glance, however, if one looks only at the data from 1946 to 1965,
then it appears very linear. Furthermore, if one looks at the 1985 plus the 2000--03 data,
then one can also draw a straight line through them. Therefore, we made individual linear
fits to the two parts of data. For the 1946--65 data, we get
$  O-C =  0.06357  + 2.32624\times 10^{-6} E  $
with $\sigma$=0.00556\,d (see solid line in Fig.~\ref{Fig:O-C}).
While for the 1985--2003 data, we have
$  O-C =  -0.00886  + 1.8807\times 10^{-6} E  $
with $\sigma$=0.00557\,d (see short dot line in Fig.~\ref{Fig:O-C}).
There is a small difference between the slopes of these two lines
but the latter intercept (corresponding to the improved initial epoch of maximum light)
differs from the former by 0.07243\,d.
If we shifted upwards the 1985--2003 data by this offset,
then we would find all the measurements lie on a straight line described by
\begin{equation}
  \begin{array}{rcl}
  O-C =  0.06383   & + & 2.32083\times 10^{-6} E .\\
       \pm0.00140  &\pm& 0.03132\nonumber
\label{eq:Lfit-shifted}
\end{array}
\end{equation}
This linear fit appears to be perfect ---
the quality of fit ($\sigma$=0.00618\,d) was improved evidently
compared to that ($\sigma$=0.01049\,d) given by Eq.(\ref{eq:Lfit}).
This line is also drawn in dash in Fig.~\ref{Fig:O-C}. It has almost the same slope
as that for the early 1946--65 data.
For these shifted data, a parabolic fit results in
\begin{equation}
  \begin{array}{rl}
  O-C =  ~~0.06378   & + ~ 2.02\times 10^{-6} E  \\
      ~~\pm0.00133   &\pm~ 0.03                 \\
 & - ~ 4.82\times 10^{-12}  E^{2}\\
 &\pm~ 1.92~~~~~~~~~
  \end{array}
\label{eq:LQfit-shifted}
\end{equation}
($\sigma$=0.0059\,d).
However, there are not adequate evidences allowing us to insert such an offset.
We recalled the maxima in 1946--65 are in $B$ band, but phase difference between
$V$ and $B$ is impossible to cause such a big offset as we can see in Fig.~\ref{Fig:curve-BV}.
This offset, if it existed, would mean that there is some sort of inconsistency in
the zero point time calibrations between the old and the new data.
That is, each observed time of maxima from 1985--2003 had a time lag equivalent to this offset
so that the resulted $O-C$ values should be shifted relative to the 1946--65 data by this offset.
Because the 1985--2003 data compose two different groups' observations collected in four years,
this kind of consistent errors in the zero point times (i.e. the starting observing times) in
each observing year are incogitable. Therefore, this possibility can be ruled out.
Another possible cause of such an offset comes from period change.
The pulsation period of the star was changing from 1965 to 1985 when
we were not looking it, but in 1985, the star went back to
exactly the same period which it had in 1946--65.
Because there was no observation in this time span, we are unable to affirm whether any changes
had occurred or not during this period.
Furthermore, there is also a time span between 1985 and 2000 without observations.
These two time spans without data bring about great uncertainties in deriving period change
through a parabolic fit to the $O-C$ residuals. Therefore we think the $O-C$ analysis
is not a suitable method for determining this star's period variability with current data.

Now it is clear that the overall profile of the $O-C$ residuals indicates a measurable period
change, but the Fourier solutions for different data sets did not resolve any period changes.
We note the fact that fewer data (42 maxima) with two vacancies of data in
1965--86 and 1985--2000 in the long investigated period (1946--2003)
does not support a reliable analysis of $O-C$.

%
%
\begin{table}
\caption{The times of maximum light of UY Cam.
Elapsed cycles ($E$) and $O-C$ based on the initial ephemeris
HJD$_{\rm max}$= 2451557.27203 + 0.26704 $E$. }
\label{Tab:O-C}
  \begin{center}
  \begin{tabular}{crrc}
\tableline\tableline
HJD(2400000+) & \multicolumn{1}{c}{$E$}  & $O-C$ & source$^{\dag}$    \\   
\tableline
32144.42300  &   -72696  &   -.10919   &   1 \\
32240.56500  &   -72336  &   -.10159   &   1 \\
32643.78700  &   -70826  &   -.10999   &   1 \\
32985.61000  &   -69546  &   -.09819   &   1 \\
33116.46600  &   -69056  &   -.09179   &   1 \\
33228.61800  &   -68636  &   -.09659   &   1 \\
33399.51900  &   -67996  &   -.10119   &   1 \\
33565.09000  &   -67376  &   -.09499   &   1 \\
33805.43700  &   -66476  &   -.08399   &   1 \\
34019.06700  &   -65676  &   -.08599   &   1 \\
34328.83900  &   -64516  &   -.08039   &   1 \\
35565.23500  &   -59886  &   -.07959   &   1 \\
36208.81400  &   -57476  &   -.06699   &   1 \\
36550.62700  &   -56196  &   -.06519   &   1 \\
36927.15600  &   -54786  &   -.06259   &   1 \\
37690.88500  &   -51926  &   -.06799   &   1 \\
38056.75300  &   -50556  &   -.04479   &   1 \\
38289.07200  &   -49686  &   -.05059   &   1 \\
38459.97900  &   -49046  &   -.04919   &   1 \\
38681.61400  &   -48216  &   -.05739   &   1 \\
38831.16800  &   -47656  &   -.04579   &   1 \\
46170.49480  &   -20172  &   -.04635   &   2 \\
46171.56400  &   -20168  &   -.04531   &   2 \\
46172.35900  &   -20165  &   -.05143   &   2 \\
46173.43020  &   -20161  &   -.04839   &   2 \\
46184.38150  &   -20120  &   -.04573   &   2 \\
51557.27203  &        0  &    .00000   &   3 \\
51560.20428  &       11  &   -.00519   &   3 \\
51561.27899  &       15  &    .00136   &   3 \\
51568.21634  &       41  &   -.00433   &   3 \\
52298.30404  &     2775  &   -.00399   &   3 \\
52301.24094  &     2786  &   -.00453   &   3 \\
52302.31347  &     2790  &   -.00016   &   3 \\
52303.12009  &     2793  &    .00534   &   3 \\
52303.37114  &     2794  &   -.01065   &   3 \\
52304.17633  &     2797  &   -.00658   &   3 \\
52306.30818  &     2805  &   -.01105   &   3 \\
52640.11870  &     4055  &   -.00053   &   3 \\
52640.38168  &     4056  &   -.00459   &   3 \\
52699.13346  &     4276  &   -.00161   &   3 \\
52701.26168  &     4284  &   -.00971   &   3 \\
52702.06761  &     4287  &   -.00490   &   3 \\
\tableline
  \end{tabular}
  \end{center}
$^{\dag}$ Note: (1) Beyer M., 1966, Astron. Nachr., 289, 95;
(2) Broglia \& Conconi, 1992, IBVS, No.3748;
(3) present data
\end{table}


\section{Discussion}
\label{sect:discuss}

\subsection{Light curves and types of variability}
An inspection on all the light curves show that they are slightly asymmetrical or
nearly sinusoidal with rounded maxima.
It was estimated that about 55\% of time ($\sim$3.5\,h) the variable be in
the descending branch.
Amplitude variations at cycle-to-cycle time-scale
did not always occur from a cycle to next. Deviation between the fit and the observations
on HJD 24512304 (2002 Jan.~29)
is unique in our 19 nights' data as well as in the 5 nights' data of \citet{brog92}.
The difference of amplitudes between 1985 and 2000s seems not to be a strong evidence for
longer time-scale amplitude variability.
We could not infer any periodicity of amplitude variation from these light curves.
On the contrary, we might interpret the deviation on 2002 Jan.~29 to be a `peculiar cycle'.
In the point of current Fourier solution, there exist other types
of variability to explain the light variations.

We know the $\delta$ Sct stars are pulsating variables with periods less than 0.3\,d,
and visual light amplitudes in the range from a few
thousands of a magnitude to about 0.8\,mag.
They occupy a position on the H--R diagram either on or
somewhat above/below the MS.
Most $\delta$ Sct stars belong to Population I, but a few variables show low
metals, low masses and high space velocities typical of Pop. II (i.e. SX Phoenicis-type stars).
The majority of the known $\delta$ Sct stars have evolved to post-MS stage.
The effective temperature range corresponds well with
the extension of the Cepheid and RR Lyrae instability strip to MS.
Pop.~I HADS having normal mass and chemical composition,
evolving away from the MS, together with SX Phe stars with low mass and metal content,
have periods of 1 to 5 hours, amplitudes of 0.3 to 0.8\,mag.
For a long time this small group (dwarf Cepheids) included with the large RR Lyr family.
They were late distinguished from RR Lyr stars by their shorter periods
and weaker absolute luminosities \citep{smith55, smith95}.
RR Lyr stars usually reside in galactic globular clusters (about 130 in our Galaxy,
e.g. IC 4499, \citealt{clem79}) of age about 10 Gyr or more
as well as in the bulge region of our Galaxy and other dwarf galaxies.
The RRc Lyr variables have lower light amplitudes ($\sim$0.5\,mag) and nearly sinusoidal
light curves with a rounded maximum. In all known cases of RRc Lyr, the dominant
mode is the radial first overtone, the periods are mostly in a range from about 0.2 to 0.5 days.

From the above, UY Cam is similar to the SX Phe stars in amplitude,
while it is similar to the RRc Lyr stars in period and in the shape and amplitude of light curves.
A high-amplitude $\delta$ Sct star
might be misclassified and could be reclassified to be SX Phe or RR Lyr-type.
To go into details of the variability types, we give further information
on the nature of the star below.

\subsection{Physical parameters}
\label{sect:phy-par}
By adopting the values $(b-y)$=0.149, $m_1$=0.110, $c_1$=1.140 and $\beta$=2.754\,mag
for UY Cam from \citet{rodr00}, we can deredden these indices making use of
the dereddening formulae and calibrations for A-type stars given by \citet{craw79}.
This way, we derive a color excess of $E(b-y)$=0.003\,mag and the
following intrinsic indices: $(b-y)_0$=0.146, $m_0$=0.111 and c$_0$=1.139\,mag.
Furthermore, deviations from the zero age main sequence
values of $\delta$$m_0$=0.077 and $\delta$c$_0$=0.452\,mag are also found.

A mean metal abundance of [Me/H]=$-$0.732\,dex (or Z=0.0037)
was derived from $\delta m_0$ using the calibrations for metallicity of A-type stars
by \citet{smal93}. A poorer value of [Me/H]=$-$1.51 was derived by
\citet{fern97}.
The metallicity is quite poorer than those of other HADS, and is comparable with those
of SX Phe-type stars, e.g. see table~2 of \citet{McNa00}.
Usually, the shorter-period variables are metal poor (SX Phe stars),
while the longer-period variables are metal strong
as seen from the table~2 and fig.~1 of \citet{McNa00}.
However, the period of UY Cam ($\log P$=$-$0.573) is the second longest (almost
equal to that of V1719 Cyg and shorter just than the period of SS Psc), but its metal abundance
derived above is so poor that the star does not fit the general
profile or relation of [Fe/H] -- $\log P$ for the high-amplitude $\delta$ Sct stars (see
fig.~1 of \citealt{McNa00}).

Stellar physical parameters of UY Cam including effective temperature,
absolute magnitude, surface gravity and other quantities can be
derived by applying suitable calibrations for the $uvby\beta$ indices.
As doing in \citet*{zhou01b, zhou02}, we derived
$T_{\rm eff}= 7300\pm150$\,K.
We used $\beta$, which is free of interstellar extinction effects,
as the independent parameter for measuring temperature \citep{craw79}.
The effective temperature was determined from
the model-atmosphere calibrations of $uvby\beta$ photometry by \citet{moon}.
Due to large value of $\delta c_0$ ($>$0.28\,mag,
too luminous relative to normal $\delta$ Sct stars)
we were misled to an unusual value of
absolute visual magnitude $M_v$=$-$1.16\,mag as well as a small surface gravity of
$\log g= 2.79\pm0.06$ by Moon's program {\sc uvbybeta} \citep{moon85}.
However, this value of $M_v$ is generally unreasonable
because it would place the star out of the $\delta$ Sct instability strip
according to several Hertzsprung-Russell (H-R) diagrams of the pulsating variables,
e.g. fig.~2 of \citet{breg00}, fig.2~of \citet{McNa00} and
fig.\ 8 of \citet{rodr01}, even though we know there are a few $\delta$ Sct stars
outside the instability strip. The gravity is also too low for a normal $\delta$ Sct star.
Therefore, we tried to determine $M_v$ and $\log g$ in other ways.
Based on $\beta$=2.754\,mag, assuming $v \sin i$=25 km s$^{-1}$ (a statistic average value
for HADS according to \citealt{jian00}), we obtained $M_v$=$-$0.228\,mag
using the relation by \citet{domi99}. This value is comparable with the results derived
from \citet{craw79} and from \citet{math92}: $-$0.454 and $-$0.218\,mag, respectively.
Furthermore,
this value has a small difference from (0.246$\pm$0.15 or 0.235$\pm$0.18\,mag) that predicted by
the period-luminosity relation for HADS, SX Phe and RR Lyr stars of
fundamental pulsation \citep{hog97, pete98, McNa97, McNa02, lane02}.
On the other hand, we can check $M_v$ from the {\it Hipparcos\/} parallax.
Unfortunately, the parallax has a great uncertainty ($-$1.64$\pm$2.08\,mas).
An examination throughout the $\delta$ Sct stars catalogue \citep{rodr00} tells us
there are only eight stars having negative parallax values with determination errors
larger than themselves (the stars are TV Lyn, CW Ser, V974 Oph, V567 Oph, BQ Ind, BP Peg,
DE Lac and UY Cam). Taking the positive value $-1.64+$2.08=0.44\,mas for UY Cam
we have $M_v$=$-$0.38\,mag if $\langle V \rangle$=11.4\,mag from the formula
$M_v = m + 5 + 5\log p$, where $p$ is parallax in parsecs.
Therefore, we tend to take $M_v$=$-$0.2$\pm$0.2 for UY Cam.
As a consequence, these parameters suggest the variable to be a
$\delta$~Sct star with poor abundance in metals and
situated into the upper $\delta$~Sct region in the H-R diagrams
as can be seen from fig.\ 8 of \citet{rodr01} and fig.2~of \citet{McNa00}.
%
%
\begin{table*}[th]
  \begin{center}
  \caption{Dereddened indices and derived physical parameters for
UY Cam.  }
  \label{Tab:phy-par}
  \smallskip
  \begin{tabular}{lclc}
\tableline\tableline
Parameter      &  Values (mag)   &   Parameter       &    Values \\
\tableline
$E(b-y)$       & 0.003$\pm$0.01  & age (Gyr)         &  0.7$\pm$0.1 \\
$(b-y)_0$      & 0.146$\pm$0.01  & $M$/M$_{\odot}$   &  2.0$\pm$0.3 \\
$m_0$          & 0.111$\pm$0.01  & $R$/R$_{\odot}$   &  5.50$\pm$0.5 \\
$c_0$          & 1.139$\pm$0.01  & $\log L/L_{\odot}$&  1.90$\pm$0.08 \\
$\delta$$m_0$  & 0.077$\pm$0.01  & $T_{\rm eff}$ (K) &  7300$\pm$150  \\
$\delta$$c_0$  & 0.452$\pm$0.01  & $\log g$ (dex)    &  3.46$\pm$0.06    \\
$M_v$          & $-0.2 \pm$0.2   & [Me/H] (dex)      & $-$0.732$\pm$0.1     \\
$M_{\rm bol}$  & 0.0$\pm$0.2   & $\bar{\rho}/\rho_{\odot}$&0.012$\pm$0.005 \\
\tableline
  \end{tabular}
  \end{center}
\end{table*}

In addition, we derived $\log g$=3.46$\pm$0.06 from the biparametric calibrations by
\citet{riba97} and the grids of $uvby$ colors for [Me/H]=$-$0.5 by \citet{smal97}.

The interstellar reddening can be neglected in terms of
the color excess of $E(b-y)$=0.003\,mag and the galactic latitude ($b$=30.8$^{\circ}$) of the star.
In fact, following \citet{craw76}
the reddening value is only about 0.0006\,mag.
Consequently, a distance modulus $(m-M_v)$ =11.2$\pm$0.2\,mag was obtained,
i.e. a distance of 1.74 Kpc by the formula 5\,$\log r = (m-M_v) + 5$.
This distance in turn results in a parallax of 0.57\,mas.
Assuming a bolometric correction, B.C.=0.02\,mag, derived from \citet{mala86}
for $T_{\rm eff}$=7300\,K,
the bolometric magnitude is $M_{\rm bol}$=0.0$\pm$0.2
($\log L/L_{\sun}$ = 1.90$\pm$0.08).
Moreover, it is possible to gain some insight into the mass and
age of this star using the evolutionary tracks of \citet{cg98}
for Z=0.004. In this case, values of an evolutionary mass
$M=2.0\pm$0.3\,M$_{\odot}$ and an age of 0.7$\pm$0.1\,Gyr are
found with $T_{\rm eff}$=7300\,K and $\log g$=3.46.
So radius $R$=5.5$\pm$0.5\,R$_{\odot}$ from radiation raw or period-radius relation
\citep{McNa78, fern92, lane02} and
a mean density of $\bar{\rho}\sim M/R^{3}$=0.012$\pm$0.005\,$\rho_{\odot}$ were derived.
Table~\ref{Tab:phy-par} tabulates the parameters derived for UY Cam.

Regarding younger age and lower surface gravity, UY Cam is similar to a Population I star.
But it could be a Pop.~II star (SX Phe-type) for its poor metal abundance and
advanced evolution stage,
or a RR Lyr-type star for its distance, period, luminosity, metallicity and
morphology of light curves.


Furthermore, with respect to long pulsating periods, \citet{rodr02} discussed the implication of
the $\delta$ Sct stars with periods longer than 0.25 days. There are only 14 stars in
this subgroup from the list of a total of 636 $\delta$ Sct stars \citep{rodr00}.
On the basis of an examination of the pulsational behaviors, luminosities,
metallicities and light curves, they reclassified three HADS, DH Peg, UY Cam and YZ Cap as
RR Lyr-type variables, while suspected three other HADS including SS Psc to be not $\delta$ Sct-type.
We found UY Cam resembles V1719 Cyg in several aspects.
However, V1719 Cyg is remained as a $\delta$ Sct pulsator.
A comparison between them was made and
listed in Table~\ref{Tab:UYCam-V1719Cyg}. It is quite interesting to conduct a further
comparison study for the two stars.

V1719 Cyg has a rotational velocity of $v\sin i$=31\,km s$^{-1}$ \citep{sola97}.
No observed $v\sin i$ value was found for UY Cam in the literature,
e.g. in the catalogue of stellar projected rotational velocities of \citet{gleb00}.
According to the statistics made by \citet{jian00} and \citet{rodr00}, as our previous estimate,
UY Cam's rotational velocity should be around 20\,km s$^{-1}$.
Observations showed that the HADS appear to be much more slowly rotating than the other
$\delta$ Sct stars, as pointed by e.g. \citet{breg00}. Statistic results of
pulsation amplitudes vs rotational velocities of known 191 $\delta$ Sct stars are that
amplitudes decrease with $v\sin i$ values increasing, i.e. high-amplitude pulsators have
low $v\sin i$ values and fewer modes, while low-amplitude pulsators are more fast rotators
with multiple simultaneously excited modes \citep{breg00,jian00,rodr00}.
This correlation between slow rotation and high amplitude may be an important clue
to the excitation and damping mechanisms for the pulsations in these stars.
However, limited by our observational conditions,
we could not obtain a spectrum for this fainter star ($V$=11.4\,mag).
If a high-resolution spectrum of this star could be taken in the future so that
some sort of model atmosphere fit could be done and then we may expect to derive
its rotational velocity, effective temperature, surface gravity, metallicity and other
atmospheric parameters.
RR Lyr variables have no detectable rotation -- \citet{pete96} estimated an upper limit
of $v\sin i <$ 10\,km~s$^{-1}$ for the RR Lyr stars. So a spectroscopic study of UY Cam
is very useful for determining the star's nature.
%
%
\begin{table*}[t]
  \begin{center}
  \caption{A comparison between two similar high-amplitude $\delta$ Sct stars
  UY Cam and V1719 Cyg. Top part lists similarities, while bottom part gives differences$^{\ddag}$.}
  \label{Tab:UYCam-V1719Cyg}
  \smallskip
  \begin{tabular}{lcc}
\tableline\tableline
        Parameter               &   UY Cam        &   V1719 Cyg \\
\tableline
1. primary period (d$^{-1}$)    &   0.26704       &   0.2673    \\
2. stable period                &    yes          &    yes      \\
3. radius (R$_{\odot}$)         &    4.9          &    5.5      \\
4. gravity, $\log g$            &    3.46         &    3.1--3.4      \\
5. $\Delta V$ (mag)             &   0.34          &    0.35     \\
6. $M_v$ (mag)                  &   $-$0.2$\pm$.2 &    0.37     \\
7. status of evolution          &  post-MS        &  post-MS    \\    
8. population                   &   I or II       &   I         \\
\hline
9. mass (M$_{\odot}$)            &   2.0$\pm$0.3   &  $\sim$2.0  \\
10. periodicity                  &   singular      &  double     \\ 
11. spectral type                &   A3-6 III      &  F5 III     \\
12. $\langle T_{\mathrm{eff}}\rangle$ (K) &   7300$\pm$150  &  6750--7300 \\
13. metallicity [Fe/H]          & poor: $-$0.732   &  rich: 0.25 \\
14. distance (pc)               &   1740$\pm$100  &  324$\pm$53 \\
15. reddening, $E(b-y)$ (mag)   &   0.02          &  0.006      \\
16. age (Gyr)                   &   0.7$\pm$0.1   &  ---           \\
\tableline
  \end{tabular}
  \end{center}
$^{\ddag}$ According to \citet{rodr02} and \citet{pena02} and references therein.
\end{table*}

SX Phe-type variables are not yet fully explained by the stellar evolution theory.
Both SX Phe and RR Lyr stars are distance scale and are very important objects for
the study of stellar evolution as well as the study of clusters and the structures of galaxies,
in which quite a number of SX Phe and RR Lyr stars inhabit. The speciality sharing part of the
photometric properties of $\delta$ Sct, SX Phe and RR Lyr variables makes UY Cam an important star.


\subsection{The mode}
From a study of phase shift and amplitude ratio between different colors' data,
UY Cam was identified as a radial pulsator \citep{rodr96}.
By means of the empirical formula of \citet{pete72},
we obtained pulsation constant $Q$=0.037$\pm$0.007\,d for $f_0$. Error might be
underestimated due to various calibrations used in deriving physical parameters.
Anyhow, this Q value means the primary frequency is a radial mode.
From the basic pulsation equation
$Q_0 = P_0 \sqrt{\bar{\rho}}$ (here $P_0$ is the fundamental period),
mean density $\bar{\rho}$=0.02\,$\rho_{\odot}$,
is about double of the value estimated above ($M/R^3\sim$0.01\,$\rho_{\odot}$).
If $\bar{\rho}$=0.01\,$\rho_{\odot}$, then Q=0.0267\,d, a value largely corresponding to first-overtone pulsation.
So $f_0$ is probably not the fundamental mode. Furthermore, owing to the uncertainty in
$M_{\rm bol}$ value, $f_0$ has possibilities to be a fundamental or first-overtone mode
according to the empirical period-luminosity-color relations \citep{stel79,lope90,tsve85}.
If $f_0$ is the first-overtone mode, we could expect the mean period fundamentalized
to be 2.913\,d$^{-1}$ by assuming normal frequency ratio $f_0/f_1$=0.778,
the ratio predicted from models. We noted there is a term at 2.98\,d$^{-1}$ appeared
in the residual spectrum after removing $f_0$ and $2f_0$,
but not significant in the last panel of Fig.~\ref{Fig:power}.
Succeeding Fourier transforms output
two frequencies at 1.364 and 2.987\,d$^{-1}$ with
amplitudes 0.018 and 0.014\,mag (below the significance level), respectively.
But they were not detected to be real at all. It is very hard to safely detect an intrinsic
oscillation frequency in this low-frequency region (0--4\,d$^{-1}$) at the above amplitude level.
If $f_0$ were fundamental mode, then the expected second frequency
would be at 4.8\,d$^{-1}$ as potential first-overtone mode.
Similarly, this term is difficult to resolve with current data.
Finally, obtaining accurate value of $M_{\rm bol}$ is very important to identify the mode
of the primary frequency. If $M_{\rm bol}$ is positive, e.g. 0.2\,mag, the mode will
be the fundamental.
In conclusion, the oscillation nature of $f_0$ is radial with possibility of
fundamental or first-overtone mode.

\section{Conclusions}
Based on the available data from 1985 to 2003, we confirm UY Cam is
a monoperiodic radial pulsator.
The light curves of UY Cam are slightly asymmetrical.
About 55\% of time ($\sim$3.5\,h) the variable be in the descending branch.

We searched the four-year data for possible frequency and amplitude variability.
Fourier analyses cannot resolve period change (see Table~\ref{Tab:freq-var}).
Forced parabolic fits to the observed minus calculated times of maximum light,
Eq.(\ref{eq:LQfit}) and (\ref{eq:LQfit-mean}), suggest
a period change rate of
$\frac{d{\rm P}}{d{\rm t}} = (-3.32\pm2.53)\times 10^{-11}$~d\,d$^{-1}$
or $(-10.0\pm2.6)\times 10^{-11}$~d\,d$^{-1}$.
We discussed the distribution of the $O-C$ data points in 1985--2003 and finally dismissed
the treatment of inserting an offset for this part of data.
Due to vacancies of data in the two periods 1965--85 and 1985--2000 and
fewer data points in the $O-C$ plot,
fits to the $O-C$ residuals suffered from great uncertainties.
The $O-C$ diagram is shown in Fig.~\ref{Fig:O-C}.
At present, we do not think the period change of UY Cam has been established
concerning the current status of data and the fitted results.
In the errors range, the amplitudes of $f_0$ were constant in 2000--2003,
but it appeared to change from 1985 to 2000s.
Table~\ref{Tab:ampl-var} shows the amplitudes in the 1985, 2000, 2002 and 2003.

According to the star's location in the color-magnitude diagram (see
fig.~8 of \citealt{rodr01}),
UY Cam is located in upper portion of the instability region of
$\delta$ Sct variables.
Using photometric indices, we derived the main physical parameters for UY Cam.
The photometric properties and stellar parameters are given in
Table~\ref{Tab:phy-par}.
UY Cam could be a Pop.~I high amplitude $\delta$ Sct star with poor metal abundances
(Z=0.0037, about 18.5 per cent of solar values)
evolving on its post-MS stage after the turn-off point (age=0.7$\pm$0.1\,Gyr)
in a shell hydrogen-burning phase.
UY Cam probably pulsates in radial first-overtone mode.

UY Cam becomes an interesting object because of its longer period, poorer metallicity,
higher luminosity, lower gravity and larger radius among the HADS.
The star intervenes in Pop.~I HADS, Pop.~II HADS or SX Phe stars and RR Lyr-type stars.
Concerning these properties and features, the variable is therefore an analogue of
both dwarf Cepheids (HADS plus SX Phe) and RR Lyr stars.

\acknowledgments
This research was funded by the National Natural Science Foundation of
China (No.10273014).

\label{lastpage}

\end{document}